\documentclass[pdflatex,sn-mathphys]{sn-jnl}

\usepackage{lineno}
\usepackage{caption}
\theoremstyle{thmstyleone}%
\theoremstyle{thmstyletwo}%

\theoremstyle{thmstylethree}%

\begin{document}

\title{Bulk-LDOS Correspondence in Topological Insulators}


\author[1,5]{\fnm{Biye} \sur{Xie}}
\equalcont{These authors contributed equally to this work.}

\author[2,3]{\fnm{Renwen} \sur{Huang}}
\equalcont{These authors contributed equally to this work.}

\author[2,3]{\fnm{Shiyin} \sur{Jia}}
\equalcont{These authors contributed equally to this work.}

\author[1]{\fnm{Zemeng} \sur{Lin}}
\equalcont{These authors contributed equally to this work.}

\author[2,3]{\fnm{Junzheng} \sur{Hu}}

\author[2,3]{\fnm{Yao} \sur{Jiang}}

\author[1]{\fnm{Shaojie} \sur{Ma}}

\author*[2,3]{\fnm{Peng} \sur{Zhan}}\email{zhanpeng@nju.edu.cn}

\author*[2,4]{\fnm{Minghui} \sur{Lu}}\email{luminghui@nju.edu.cn}

\author[2,3]{\fnm{Zhenlin} \sur{Wang}}

\author[2,4]{\fnm{Yanfeng} \sur{Chen}}

\author*[1]{\fnm{Shuang} \sur{Zhang}}\email{shuzhang@hku.hk}

\affil*[1]{\orgdiv{Department of Physics}, \orgname{The University of Hong Kong}, \orgaddress{\street{Pokfulam Road}, \city{Hong Kong}, \country{China}}}

\affil[2]{\orgdiv{National Laboratory of Solid State Microstructures, Collaborative Innovation Center of Advanced Microstructures}, \orgname{Nanjing University}, \orgaddress{\city{Nanjing}, \postcode{518172}, \country{China}}}

\affil[3]{\orgdiv{School of Physics}, \orgname{Nanjing University}, \orgaddress{\city{Nanjing}, \postcode{518172}, \country{China}}}

\affil[4]{\orgdiv{Department of Materials Science and Engineering}, \orgname{Nanjing University}, \orgaddress{\city{Nanjing}, \postcode{210093}, \country{China}}}

\affil[5]{\orgdiv{School of Science and Engineering}, \orgname{The Chinese University of Hong Kong, Shenzhen}, \orgaddress{\postcode{518172}, \country{China}}}

     \abstract{Seeking the criterion for diagnosing topological phases in real materials has been one of the major tasks in topological physics~\cite{RMP_Zhang_Qi,RMP_Kane}. Currently, bulk-boundary correspondence based on spectral measurements of in-gap topological boundary states~\cite{WangZheng_Nature,HuXiao_prl} and the fractional corner anomaly derived from the measurement of the fractional spectral charge~\cite{LDOS1_FCA} are two main approaches to characterize topologically insulating phases. However, these two methods require a complete band-gap with either in-gap states or strict spatial symmetry of the overall sample which significantly limits their applications to more generalized cases~\cite{WangZheng_Nature,HuXiao_prl,LDOS1_FCA}. Here we propose and demonstrate an approach to link the non-trivial hierarchical bulk topology~\cite{Ourprl} to the multidimensional partition of local-density of states (LDOS) respectively ~\cite{LDOS2_quadrupole_Huges,LDOS1_FCA,LDOS3_disclination_Huges,disclination_Jianhua_Jiang}, denoted as the bulk-LDOS correspondence. Specifically, in a finite-size topologically nontrivial photonic crystal, we observe that the distribution of LDOS is divided into three partitioned regions of the sample - the two-dimensional interior bulk area (avoiding edge and corner areas), one-dimensional edge region (avoiding the corner area) and zero-dimensional corner sites. In contrast, the LDOS is distributed across the entire two-dimensional bulk area across the whole spectrum for the topologically trivial cases. Moreover, we present the universality of this criterion by validate this correspondence in both a higher-order topological insulator without a complete band-gap and with disorders~\cite{Anderson_Alexander_Rechtsman_Segev,Anderson_Huges,Fleury}. Our findings provide a general way to distinguishing topological insulators and unveil the unexplored features of topological directional band-gap materials without in-gap states.}

\maketitle

\section*{Introduction}\label{sec1}

Topological materials which go beyond the spontaneous symmetry breaking paradigm have expanded our understanding of condensed matter physics~\cite{RMP_Topological_photonics,tp_mater1,tp_mater2,tp_mater3,tp_mater4} and show promising applications in more energy-efficient electronic devices~\cite{gilbert2021topological} and quantum computing~\cite{nayak2008non} ascribed to their robust and unique transport properties~\cite{xiao2015geometric,huang2011dirac,laser1,laser2,ap1,ap2,ap3,ap4}. Theoretically, each topological phase of matter can be characterized by a quantized number of bulk physics, denoted as the topological invariant~\cite{kane2005z,fu2007topological}. However, in experiments, the direct measurement of bulk topological invariants is quite difficult since it needs sophisticated quantum states tomography~\cite{li2016bloch}. Fortunately, correspondences exists between bulk topological physics and other experimental observables. For example, for 
completely gapped topological phase (or topological insulators, TIs), conventional bulk-boundary correspondence (BBC)~\cite{chiu2016classification} dictates the existence of in-gap topological boundary states at the interface between two topologically distinct materials. The powerful BBC is one of the most important features of TIs and has been generalized to higher-order topology~\cite{BBC_Higher_order}, non-Hermitian topology~\cite{BBC_nonH}, and 4D quantum Hall systems~\cite{wang2020circuit}. However, in a series of recent studies of topological crystalline insulators~\cite{TCI_FuLiang1,TCI_FuLiang2,benalcazar_Science_multipole,Huges_Chiral2018,benalcazar2019prb,disclination_Jianhua_Jiang}, topological phases without chiral symmetry (or particle-hole symmetry) can host boundary states that are embedded into the bulk spectrum and thus the BBC fails to precisely distinguishing this kind of topological phases. 

    To overcome this problem, Peterson, et al~\cite{LDOS1_FCA} put forward a measurable topological indicator to identify non-trivial higher-order topological crystalline phases without the requirement of in-gap localized boundary states. Specifically, they measure the portion of Wannier centers (WCs) (also called spectral charges) by integrating the local density of state (LDOS) in each unit cell at bulk, edge and corner area respectively for entire band and then defined a fractional corner anomaly (FCA) which links the non-trivial higher-order topological insulating phases to an observable fractional quantum number. Lately, based on similar measurement of the fractional quantum number (related to the number of WCs in a unit cell), a bulk-disclination correspondence has been established to predict the existence of topological disclination states with various spatial symmetries~\cite{LDOS3_disclination_Huges,disclination_Jianhua_Jiang}. But there still exists a fundamental limitation for this correspondence: the WCs can only be defined (and thus measured) over an entire band (or bands) and their portions in unit cells are heavily influenced by lattice spatial symmetries. For a large family of TIs with directional bandgaps (also called partial bandgaps)~\cite{partial1} and no in-gap states, or even with disorders that breaks all spatial symmetries, both the BBC and FCA fail to characterize topological phases. Hence a crucial question arises: what is the general correspondence between non-trivial bulk topology and measurable observables?

In this work, we propose and demonstrate an approach to precisely identify distinct topological phases charactered by WCs by observing the multidimensional and single-dimensional partition of LDOS for topologically nontrivial and trivial phases respectively and hence provide a general correspondence between non-trivial bulk topology and measurable LDOS. To overcome the finite-size effect, we further obtain the averaged LDOS over a small energy domain and over the bulk, edge and corner areas for both topologically non-trivial and trivial lattices, we find that the magnitudes of the LDOS averaged over these area are almost identical regardless of the energy magnitude for trivial phases while they are significantly 
different for nontrivial phases. We provide an intuitive explanation of this effect based on modern polarization theory~\cite{benalcazar2019prb} and experimental demonstration in a two-dimensional (2D) photonic higher-order TIs with directional bandgaps and no in-gap localized boundary states. Finally, we validate this correspondence in systems even with random disorders. Our finding provides a general criterion to diagnose topological phases and offers possibilities to study topological phases in topological directional bandgap materials.

\section*{Multidimensional partition of LDOS reveals topology}

To have an intuitive understanding of this approach, we start from a TI where the topological phases are characterized by the displacement of WCs in unit cells (bulk polarizations and filling anomaly)~\cite{ezawa2018higher,xue2019acoustic,ni2019observation,benalcazar2019prb}. For TIs, all WCs located away from centers of unit cells. Specifically, WCs located at the sides and corners of unit cells corresponds to the first-order and second-order TIs respectively as shown in Fig. 1\textbf{a}. On the contrary, if WCs are located at the centers of unit cells, the system is a trivial atomic topological insulator (or ordinary insulator, OI) (see Fig. 1\textbf{b}). Since WCs represent charge centers of wavefunctions~\cite{marzari2012maximally}, the number of WCs in certain areas determines the magnitude of LDOS in those areas. Moreover, in a finite-size lattice, WCs located at bulk, edge and corner areas (see colored areas in Fig. 1) determine the portion of waves that hybridize to form bulk, edge and corner states respectively. As a consequence, for topologically nontrivial lattice with WCs localized at the corner of unit cells as shown in Fig. 1a, the multidimensional hybridization of single lattice site's orbitals at bulk(grey area in Fig. 1a), edge(blue area in Fig. 1a), and corner areas (red area in Fig. 1a) leads to the multidimensional partition of LDOS in the spectrum. However, for topologically trivial lattice, all lattice sites' orbitals (represented by grey area in Fig. 1b) hybridize to form bulk states and therefore the LDOS is distributed in single-dimension (here is 2-dimension for example). This multidimensional partition of LDOS in a system with open boundary condition is an intrinsic physical properties of topologically non-trivial phase and it recovers to the conventional BBC when there is a complete bandgap with in-gap localized boundary states and FCA when there is a complete bandgap with crystalline symmetries~\cite{LDOS1_FCA}. Moreover it can characterized topologically non-trivial phases even in systems with directional bandgap and no in-gap boundary states or without any symmetries. The displacement of WCs in unit cells which is the bulk topological invariant has one-to-one correspondence to the multidimensional partition of LDOS in a system with open boundary condition, clearly revealing an intrinsic bulk-LDOS correspondence of topological phases (see detailed discussions in Section I in SI~\cite{SI}).

Although we can directly apply the bulk-LDOS correspondence to distinguish different topological phases, in a finite-size structure with a few unit cells, it is not always clear to see the multidimensional partition of LDOS at an arbitrary energy magnitude due to the discrete resonating mode in the whole structure. To overcome this problem, we note that in a finte-size structure with open boundary condition, different-dimensional hybridized states have different consecutive level spacings (the frequency distances between frequency-adjecent modes)~\cite{scaling}, therefore the bulk, edge and corner states are inhomogeneously distributed in the energy domain (see the dots in the vertical axis in Fig. 1\textbf{c}) for TIs. In terms of the LDOS, the running average of the LDOS for eigenmodes over a small energy range are also distributed in different dimensions in the energy domain (see Fig. 1\textbf{c}). This running average of the LDOS over a small energy range will reduce the finite-size fluctuations of LDOS for a single frequency mode while preserving the character of the multidimensional partition of LDOS.  However, for OIs, the running average of the LDOS over a small energy range are single-dimensionally distributed instead (see Fig. 1\textbf{d}).

\section*{Bulk-LDOS correspondence in topological directional bandgap materials}

To experimentally observe this correspondence, we consider a 2D topological photonic crystal (PC), consisting of dielectric cylinders. PCs with engineered photon band structures have been used to explore various topological phases including quantum Hall states~\cite{wang2009observation}, quantum spin Hall states~\cite{wu2015scheme}, higher-order topological insulators~\cite{xie2019visualization}, and with fractional quantum numbers~\cite{disclination_Jianhua_Jiang}. The WCs and spectral charges can also be obtained in photonic bands. Previous studies show that by retreiving the $S_{11}$ parameters and meanwhile considering the Purcell effect in the near field scanning  process, one can directly obtain the LDOS of photonic states (see detailed discussion in Method). We here consider a 2D photonic SSH model with a directional bandgap in the $s$-wave band structure (see band structure in SI). By adjusting the intercell coupling strength $t_{inter}$ and the intracell coupling strength $t_{intra}$, the 2D photonic SSH model experiences a topological phase ($t_{inter}>t_{intra}$) or a trivial phase ($t_{inter}<t_{intra}$). Moreover, the 2D photonic SSH model has been previously shown to have hierarchical topological phases with both 1D edge states and 0D corner states~\cite{xie2019visualization}. The higher-order topological phase in this photonic crystal can be numerically classified by the displacement of WCs of photonic modes and explained by the filling anomaly. To observe the 1D edge states, we place an excitation source at the 1D interface between topologically nontrivial configuration and trivial configuration of the PC (see Fig. 2\textbf{a}). The projected band structure with in-gap 1D edge states can be see from the Fourier transformation of the excited edge states (see Fig. 2\textbf{b}). We then modify boundary sites by reducing the diameters of rods (see Fig. 2\textbf{a}). A small perturbation on the onsite energy of the boundary sites will not destroy the topological phases but only shift the frequency of edge states. Consequently, the 1D topological edge states are now embedded into the bulk spectrum as shown in Fig. 2\textbf{c} and there is no in-gap states in the directional bandgaps.

Now we fabricate a 2D sample with the same parameters as those in Fig. 2 to investigate the hierarchical topological phases (The detailed sample parameters is discussed in Method) as shown in Fig. 3\textbf{a}. We then apply the near field scanning method by positioning a metal probe near the top of the sample and extracting the $S_{11}$ parameters to obtain the LDOS of photonic states (see discussions on the measurement of photonic LDOS in Method). We consider a structure of photonic crystals with modified boundary rods, perfect electric conduction boundary condition, and $6\times6$ unit cells (see Fig. 3\textbf{a}). The topological phases and bandgaps of this structure can be adjusted by the inter-cell couplings and intra-cell couplings which are determined by the distances between rods. Specifically, we realize photonic OI and TI with directional bandgap and topological boundary states embedded in the bulk spectrum similar as those in Fig. 2 (see detailed discussion in SI). The detailed design of these photonic crystals is discussed in Method. According to Fig. 2, there are no complete bandgaps from the spectra of eigenmodes and therefore one can not apply the direct measurement of spectral charges to distinguish topological phases.  

Nevertheless, by applying the same approach as discussed in previous section, we can experimentally obtain the multidimensional partition of photonic LDOS as shown in Fig. 3\textbf{b}-\textbf{c} for both non-trivial and trivial configurations. To provide a clear characterization of this multidimensional partition of the LDOS, we define a ratio $r$ between the averaged LDOS over a certain energy range $\Delta f$ starting from an initial frequency $f_i$ at edge (corner) sites $D^{i}_{edge}$ ($D^{i}_{corner}$) and those at the bulk sites $D^{i}_{bulk}$ as $r^{(n,m,i)}=\frac{D^{i}_{n}}{D^{i}_{m}}$. Here ${n,m}=bulk, edge, corner$. We set $\Delta f=1/8\Sigma$ where $\Sigma=f_t-f_b$ represents the energy range from the bottom of the band structure $f_b$ to the top of the band structure $f_t$. Here this running averaging range $\Delta f$ is chosen to ensure that the finite-size fluctuation of LDOS is reduced. We plot $r^{(edge, bulk)}$, $r^{(corner, bulk)}$, and $r^{(corner, edge)}$ with $f_i$ starting from $f_b$ to $f_t-\Delta f$ as shown in Fig. 3\textbf{b} respectively for both TIs and OIs.

We clearly see two peaks of $r^{(edge, bulk)}$ and one peak of $r^{(corner, bulk)}$ and $r^{(corner, edge)}$ for the TI and small fluctuations of $r^{(edge, bulk)}$, $r^{(corner, bulk)}$ and $r^{(corner, edge)}$ for the OI. Even though there are higher-order couplings in PCs compared to the tight-binding model and their band structures are different in the low frequency regime, the multidimensional partition of running averaged photonic LDOS is preserved and hence it captures different topological phases well. From the single frequency LDOS as shown in Fig. 3c, we find that for OI (the upper panels in Fig. 3c), when we increase the frequency from the bottom to the top of the band structure, the LDOS is distributed across the 2D entire bulk area (EBA) without any partition. However, for the TI, we observe that the LDOS is distributed across the 2D interior bulk area (IBA, avoiding 1D edges and 0D corner sites) at $4.04$GHz, both 2D IBA and 1D edges (avoiding 0D corners) at 5.04GHz, both 2D IBA and 0D corners (avoiding 1D edges) at 5.88GHz, and both 2D IBA and 1D edges (avoiding 0D corners) at 7.08GHz and 7.52GHz as shown in lower panels in Fig. 3c. This character clearly reveals the multi-dimensional partition of the LDOS for the topologically nontrivial phase.

\section*{Bulk-LDOS correspondence in disordered topological materials}

The Bulk-LDOS correspondence roots in the redistribution of the positions of WCs when there is a topological phase transition. Global spatial symmetries and chiral symmetry only determine the fractionalization of charges and the frequency of corner states respectively. A small perturbation that does not close the band gaps will not destroy the relevant topological phases. Hence, our proposed bulk-LDOS correspondence also holds for topological phases with certain level of random disorders where one cannot define the band structures and the symmetry-protected fractionalization of spectral charges. To unveil this correspondence, we add a random perturbation of on-site energy by randomly changing diameters of the rods in the unperturbed lattice (with the same parameters as those in Fig. 2b). For the fabrication simplicity, we add the random perturbation on diameters of rods in a discrete way. Specifically, based on the finite-size structure as discussed in Fig. 3\textbf{a} without modifying the boundary sites, we change diameters of rods by a number $\lambda$ which is randomly and discretely distributed between $0$ and $1/6r_0$ where $r_0=0.11a$ is the radius of the unperturbed rods. Under the small perturbations, there is still no complete bandgap in the spectra. Similarly as before, we can measure the multidimensional partition of running averaged LDOS $r^{(edge, bulk)}$, $r^{(corner, bulk)}$ and $r^{(corner, edge)}$ as shown Fig. 4\textbf{b}. We find there are two peaks for $r^{(edge, bulk)}$, indicating the first-order topology with two non-degenerate edge states and one peak for $r^{(corner, bulk)}$ and $r^{(corner, edge)}$, indicating the second-order topology with corner states. We can also see the bulk-LDOS correspondence from the single frequency measurement of LDOS for the disordered structures as shown in Fig. 4\textbf{c}. The upper and lower panels in Fig. 4\textbf{c} corresponds to the LDOS of wave functions for OI and TI with disorders respectively. We also see single-dimensional partition of LDOS for OI and a multidimensional partition of LDOS for TI. We here note that due to the random disorders in the structure, the LDOS of single mode may have no $C_4$ symmetry as shown in Fig. 4\textbf{c} and therefore the quantized fractional charges do not exist~\cite{LDOS1_FCA}. Nevertheless, the bulk-LDOS correspondence holds on. Besides, this perturbation will also break the bound-state in the continuum (BIC) and therefore one cannot use an excited source to map out the eigenfield distributions. The excited field distributions may change significantly with respect to different positions of the source and therefore cannot be applied to diagonse the bulk topology (see detailed discussions in SI~\cite{SI}).

\section*{Conclusion}

    In conclusion, we have proposed and demonstrated a rigorous Bulk-LDOS correspondence in TIs in which the topological phases are characterized by the displacement of WCs. Our approach based on the measurement of the multidimensional partition of LDOS can identify both the first-order topological insulating phases and higher-order topological phases simultaneously. Moreover, compared to previous BBC and FCA approach, the proposed bulk-LDOS correspondence can be applied to more general cases such as the TIs where there is a directional bandgap with no in-gap states and even with disorders. Although our experiment is based on photonic crystals, our findings are general and can be extended to other systems such as electronic materials~\cite{graphdiyne}, acoustic crystals~\cite{acoustic}, mechanics~\cite{huber2016topological} and plasmonics~\cite{jin2017infrared}. Besides, we expect further exploration of this correspondence in Wannier-nonrepresentable TIs such as fragile topological phases~\cite{song2020twisted} and topological phase characterized by a non-zero Chern number~\cite{wang2020circuit} and even gapless topological systems such as the Weyl and Dirac semimetals~\cite{fu2022quantum}. In applications, since the LDOS have been previously shown to be proportional to the spontaneous emission rate of atoms and molecules embedded in a photonic crystal~\cite{emission}, our findings will shed light on future explorations of topological light emissions and light-matter interactions in directional topological bandgap materials.

\section*{Methods}\label{method}

\subsection*{Design of samples}

We fabricate our photonic crystals by using $Z_r O_2$ rods with a dielectric constant $\epsilon=28$. The height of the cylinders is $9.5$mm for all samples. The lattice constant of the unit cell is $20$mm. Each sample is made of an aluminum plate with dielectric pillars inserted into shallow holes of $0.5$mm depth for fixation. A thin film of air outside the PC with a thickness of $0.125\times a$ is introduced between the PCs and the aluminum frames(serve as perfect electric conduction boundaries) on each side of samples, as depicted in Extended data Fig. 1.

\subsection*{Measurement of projected band structure}

The projected energy band is measured through a $2$-port transmission spectrum $S_{21}$ using a microwave network analyzer Keysight E5063A. We put the sample on a displacement stage with a step size of $2$mm and place a metallic plate right above the sample to prevent radiative loss. The airgap between the sample and the upper plate is $2$mm. Two antennas are included in this measurement: one is fixed through the sample to work as an excitation source and the other is embedded in the upper plate with a distance to the PCs of $1$mm for detection. Therefore, both the amplitude and phase corresponding to the eigenmode at different positions of the sample can be measured with the moving displacement stage. During each moving step, a pause of $0.5$ second is set for stationary measurement. 

\subsection*{Measurement of LDOS}

The LDOS of the PCs can be derived from the one-port reflection spectrum $S_{11}$, which is accomplished by inserting an antenna into the upper metal plate with a coaxial cable of $50\Omega$ connected to the analyzer. Between the sample and the upper metallic plate, an airgap of 2mm is introduced. We averagely divide each unit cell to $10\times10$ parts and detect the central region of each part in a range of $2-9$GHz with a resolution about $5$MHz (see Extended Data Fig. 2). 

We first calculate the extinction rate from the reflection as $E(f)=1-\|S_{11} (f)\|^2$. Considering the Purcell effect, we divide the measured extinction rate by the frequency squared, $D(f)=\frac{E(f)}{f^2}$, which is proportional to the density of states with each measured spacial point. To obtain the LDOS for one dielectric pillar $D(\textbf{r}, f)$, we sum the $D(\textbf{r}, f)$ in one quarter of the unit cell up and normalize it as:
\begin{equation}
D(\textbf{r}, f)=\sum_i D_i(\textbf{r}, f) \sigma_i
\end{equation}
with $\int_{all}D(\textbf{r}, f)=1$
Here,the integral is taken over all bands. $D(\textbf{r}, f)$ and $\sigma_i$ are the density of states and the area of the $i$th region with $i$ varying from 1 to 25 respectively (see Extended data Fig. 2). 

To demonstrate the inhomogeneous distribution of LDOS, we average $D(\textbf{r}, f)$ of each pillar belonging to corner, edge and bulk, respectively. Then, the obtained $D_{corner}(f)$, $D_{edge}(f)$ and $D_{bulk }(f)$ are further integrated over the range of $\Delta f$ starting from an initial frequency $f_i$. 
\begin{equation}
D^i_m=\int^{f_i+\Delta f}_{f} D_m(f) df
\end{equation}
Here $m=corner, edge, bulk$. $f_i$ goes from the bottom frequency of the band structure $f_b$  to the top of the band structure $f_t$. We here choose $\Delta f=1/8(f_t-f_b)$ for the reason that this value is large enough to reduce the fluctuations of averaged LDOS due to the finite-size effect and small enough to ensure the summed set of eigenmodes are firstly all bulk states and then contain the edge states as $f_i$ increases so that we can observe the spectral  inhomogeneity more clearly.  For each $f_i$, we calculate the ratio as 

\begin{equation}
r^{(n,m,i)}=\frac{D^i_m}{D^i_n}
\end{equation}
Here 
$m, n=corner, edge, bulk$. Specifically, we calculate $r^{(edge, bulk)}$, $r^{(corner, bulk)}$ and $r^{(corner, edge)}$ as shown in Fig. 3 and Fig. 4.

\subsection*{Simulation}
Numerical simulations are performed via the commercial finite-element simulation software(COMSOL MUTIPHYSICS). We build $3$D photonic crystal models in all of the simulations for better correspondence with the experiments. Energy bands with infinite structures are calculated by setting the boundaries perpendicular to the propagation direction of the edge states as periodic boundary and other boundaries as Perfect Electric Conductor(PEC). For the calculation of the LDOS with finite structure, we set PEC boundary conditions in all directions.

\section*{Data availability}
The data that support the findings of this study are available from the corresponding author upon request. Source data are provided with this paper.

\bmhead{Acknowledgments}
We acknowledge useful discussions with Oubo You. This work was financially supported by the Hong Kong Research Grant Council (AoE/P-701/20, 17309021), National Key R\&D Program of China (Grants No. 2018YFA0306200), National Natural Science Foundation of China (Grants No. 12174189 and 11834007), and the startup funding of the Chinese University of Hong Kong, Shenzhen (UDF01002563).

\bmhead{Author contributions}
B.X., P.Z., M.H.L., and S.Z. conceived the idea. B.X. and Z.M.L. performed the theoretical anlysis. Z.L., R.H., and S.J. did the numerical analysis. R.H., and S.J. did the experiment. P.Z., M.H.L., and S.Z. guided the project. All authors contributed to the discussion and writing of the manuscript.
\bmhead{Competing interests}
The authors declare no competing interests.

\bibliography{sn-bibliography}

\pagebreak
\begin{figure}[h]%
\centering
\includegraphics[width=1\textwidth]{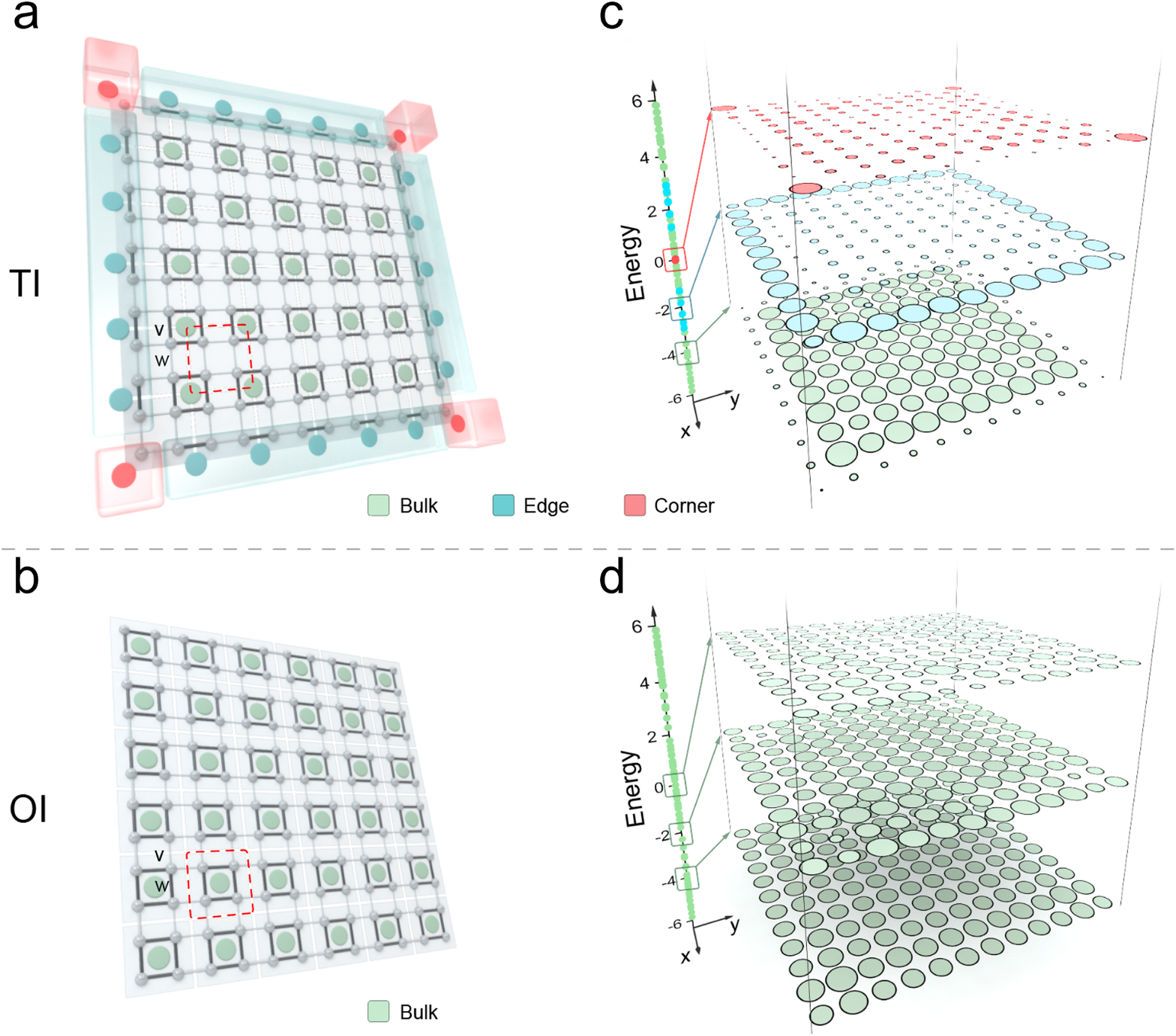}
\caption{\textbf{Schematic of the bulk-LDOS correspondence.} \textbf{a} a topological insulator (TI) with Wannier centers (WCs) located at the corner of unit cells. The grey dots, grey lines and black lines represent the lattice sites, intra-cell couplings and inter-cell couplings respectively. The green, blue and red solid circles represent WCs which are located at the bulk area (grey color area), edge area (blue color area) and corner areas (red color area) respectively. \textbf{b} a trivial atomic insulator (or ordinary insulator, OI) with WCs located at centers of unit cells. The elements have the same meaning in \textbf{a}. \textbf{c}-\textbf{d} The running averaged local density of states (LDOS) distributions over a small energy range at different positions in the energy domain for a TI as shown in \textbf{a} and OI as shown in \textbf{b} respectively. The green, blue and red dots on the axis represent the eigenstates which have a higher field strength on the bulk, edge and corner lattice sites respectively. The sizes of green, blue and red circles represent the magnitude of running averaged LDOS over the eigenstates inside the green, blue and red boxes at each lattice sites respectively. }\label{fig1}
\end{figure}

\pagebreak
\begin{figure}[h]%
\centering
\includegraphics[width=1.0\textwidth]{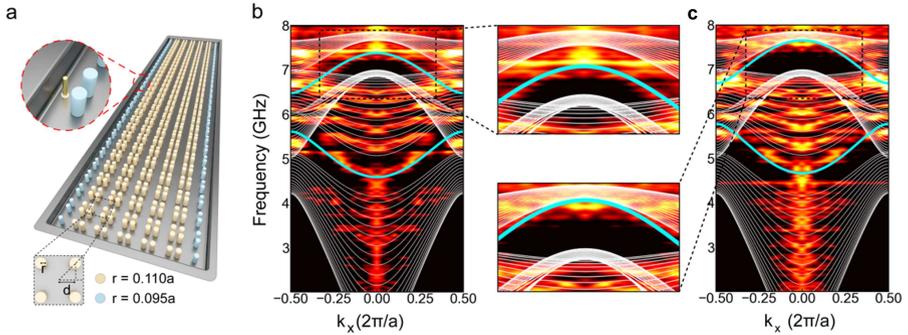}
\caption{\textbf{The projected band structures (PBS) of TIs with directional bandgaps.} \textbf{a} Schematic of a finite-size photonic TI with boundary rods having smaller radius than the bulk rods. The zoom-in inset present the position of probe to measure the projective band structure. The radius of the bulk and edge rods are $0.11a$ and $0.1$a respectively. \textbf{b} The PBS of unperturbed lattice. The white (blue) lines are the numerically simulated bulk and edge states. The bright and dark color represent the experimentally measured PBS. The zoom-in inset shown the 1D edge state is a in-gap state. \textbf{c} The PBS of perturbed lattice. The elements have the same meaning as in \textbf{b}. The zoom-in inset shown the 1D edge state is embedded into the bulk spectrum for the second band gap.}\label{fig2}
\end{figure}

\pagebreak
\begin{figure}[h]%
\centering
\includegraphics[width=1.0\textwidth]{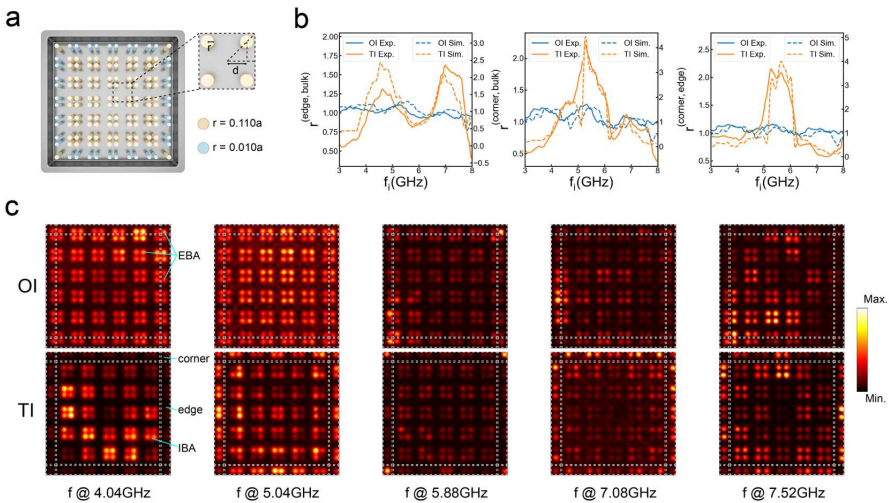}
\caption{\textbf{Multidimensional partition of photonic LDOS of a TI with perturbed boundary.} \textbf{a} A finite-size photonic TI. The parameters are the same as those in Fig. 2\textbf{a}.\textbf{b} The ratio between the running averaged edge LDOS over the runnning averaged bulk LDOS (left panel), the ratio between the runnning averaged corner LDOS over the running averaged bulk LDOS (left panel), and the ratio between the running averaged corner LDOS over the running averaged edge LDOS (left panel) are presented respectively. The solid (dashed) blue and orange lines represent measured (simulated) ratio for the OI and TI respectively. \textbf{c} The upper (lower) panels show the LDOS of a single frequency at $4.04$GHz, $5.04$GHz, $5.88$GHz, $7.08$GHz, and $7.52$GHz for OI and TI respectively. The finite-size sample is denoted as the entire bulk area (EBA) which can be divided into three separated parts (encircled by dashed lines): the interior bulk area (IBA), the 1D edges and the 0D corners.}\label{fig3}
\end{figure}

\pagebreak
\begin{figure}[h]%
\centering
\includegraphics[width=1.0\textwidth]{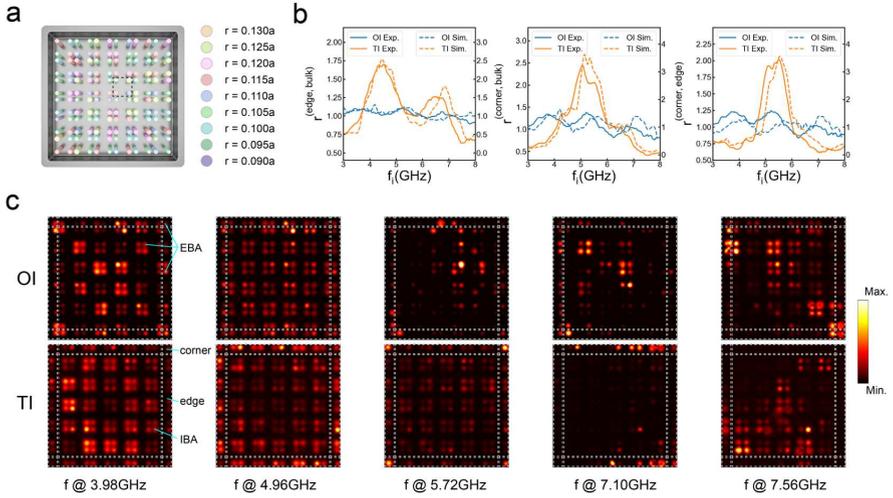}
\caption{\textbf{Multidimensional partition of photonic LDOS of a TI with random disorders.} \textbf{a} A photonic TI with discretized random disorders on the radius of the rods. The rods with the same color have the same radius. The specific value of the radius of rods are presented. \textbf{b} The ratio between the running averaged edge LDOS over the runnning averaged bulk LDOS (left panel), the ratio between the runnning averaged corner LDOS over the running averaged bulk LDOS (left panel), and the ratio between the running averaged corner LDOS over the running averaged edge LDOS (left panel) are presented respectively. The solid (dashed) blue and orange lines represent measured (simulated) ratio for the disordered OI and TI respectively. \textbf{c} The upper (lower) panels show the LDOS of a single mode at $f_i=3.98$GHz, $4.96$GHz, $5.72$GHz, $7.10$GHz,and $7.56$GHz for disordered OI and TI respectively.}\label{fig4}
\end{figure}

\pagebreak
\begin{figure}[h]%
\centering
\includegraphics[width=1.0\textwidth]{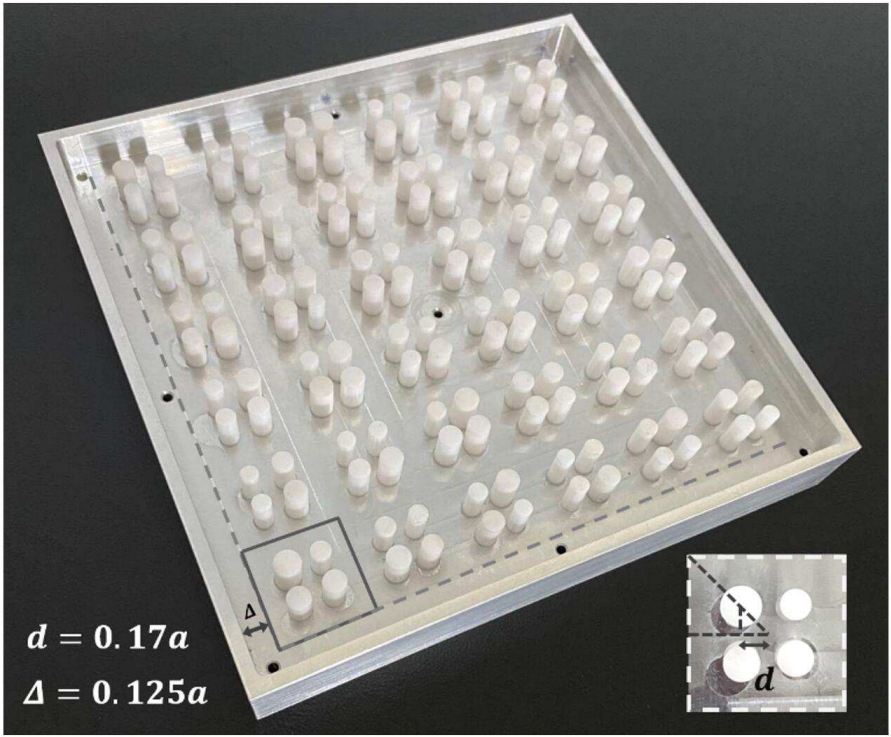}
\caption*{\textbf{Extended Data Fig. 1. Photograph of the sample.} The boundary of PCs and the air gap is indicated as the dash line, $\Delta=0.125a$ is the thickness of the air gap. The lower right panel is a zoom-in picture of the unit cell where $d$ denotes the distances of the dielectric cylinders from the center of the unit cell both horizontally and vertically. The picture shown here is the sample of $d=0.17a$ with onsite potential disorder.}
\end{figure}

\pagebreak
\begin{figure}[h]%
\centering
\includegraphics[width=1.0\textwidth]{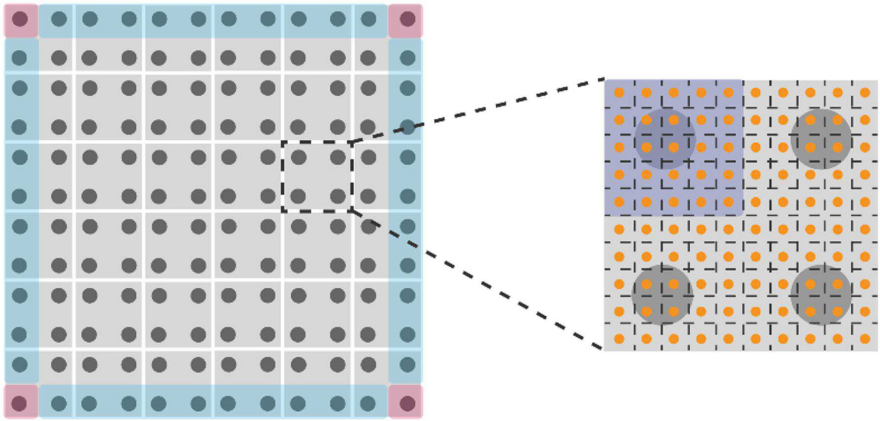}
\caption*{\textbf{Extended Data Fig. 2. Schematic of sampled points.} Unit cells and the dielectric cylinders are represented as gray squares and darker circles. The blue and pink regions denote the edges and corners of the sample, respectively. Each unit cell is divided into $10\times10$ equivalent areas with orange dots locate at the center of each area which are the measured points. }
\end{figure}

\end{document}